\documentclass[]{aa} % for a referee version
\usepackage{graphicx}
\usepackage{natbib}
\usepackage{txfonts}
\def\halpha{\mbox{H$\alpha$}}

\def\lya{\mbox{Ly$\alpha$}}

\def\lesssim{\mathrel{\hbox{\rlap{\hbox{\lower4pt\hbox{$\sim$}}}\hbox{$<$}}}}
\def\gtrsim{\mathrel{\hbox{\rlap{\hbox{\lower4pt\hbox{$\sim$}}}\hbox{$>$}}}}

\begin{document}
   \title{Uncovering strong \ion{Mg}{ii} absorbing galaxies}
   \subtitle{Imaging below the Lyman limit 
  \thanks{Based on observations collected at
 the European Southern Observatory, Chile, under programme IDs 380.A-0350 and
        080.A-0482}}
   \author{L. Christensen
          \inst{1}
          \and P. Noterdaeme
          \inst{2}
          \and P. Petitjean
          \inst{3}
          \and C. Ledoux 
          \inst{4}
          \and J. P. U. Fynbo 
          \inst{5}
%      \fnmsep\thanks{show the usage of the elements in the author field}
          }
   \institute{European Southern Observatory,
     Karl-Schwarzschild-Strasse 2, D-85748 Garching, Germany  
     \email{lichrist@eso.org}
         \and
     Inter-University Centre for Astronomy and Astrophysics, Post Bag
     4, Ganeshkhind, 411007 Pune, India - \email{pasquiern@iucaa.ernet.in}
         \and
     UPMC Paris 6, Institut d'Astrophysique de Paris,
      UMR7095 CNRS, 98bis Bd Arago, 75014 Paris, France,
     \email{petitjean@iap.fr}
         \and
     European Southern Observatory, Alonso de C\'ordova 3107, Casilla
     19001, Vitacura, Santiago 19, Chile,  \email{cledoux@eso.org}
         \and
     Dark Cosmology Centre, Niels Bohr Institute, University of 
     Copenhagen, Juliane Maries Vej 30, 2100 Copenhagen {\O}, Denmark,
     \email{jfynbo@astro.ku.dk}
%             \thanks{The university of heaven temporarily does not
%                     accept e-mails}
             }

   \date{Received 10 March 2009 / Accepted 8 August 2009}

  \abstract
  % context heading (optional) {} 
{The nature of the galaxies that give rise to absorption lines, such
  as damped Lyman-$\alpha$ systems (DLAs) or strong \ion{Mg}{ii} lines, in
  quasar spectra is difficult to investigate in emission. These
  galaxies can be very faint and located close to the lines of sight
  of the much brighter background quasars.}
  % aims heading (mandatory) 
{Taking advantage of the total absorption of the QSO light bluewards
  of the Lyman limit of two DLAs at $z>3.4$, we look for the
  continuum emission from intervening galaxies at $z\approx2$ that are
  identified via strong metal absorption lines. The \ion{Mg}{ii}
  absorbers have equivalent width large enough to be potential DLA systems. }
  % methods heading (mandatory)  
 {Deep images are obtained with the FOcal Reducer and Spectrograph
   (FORS1) on the Very Large Telescope for the fields towards
   SDSS~J110855+120953 and SDSS~J140850+020522. These quasars have
   \ion{Mg}{ii} absorption lines at $z=1.87$
   ($W_{\mathrm{r}}$(\ion{Mg}{ii})~=~2.46~{\AA}) and $z=1.98$
   ($W_{\mathrm{r}}$(\ion{Mg}{ii})~=~1.89~{\AA}), respectively,  and
     each QSO has two intervening higher redshift DLAs at
     $z>3$.  The $U$ and $R$ bands of
   FORS1 lie blue and redwards of the Lyman limit of the background
   DLAs, allowing us to search for emission from the foreground galaxies
   directly along the lines of sight to the QSOs.}
  % results heading (mandatory)
{No galaxies are found close to the sight line of the QSO to a point
  source limit of $U_{\mathrm{AB}}\sim28.0$. In both fields, the
  closest objects lie at an impact parameter of
  $\sim$5\arcsec\ corresponding to $\sim$40~kpc in projection at
  $z=2$, and have typical colours of star forming galaxies at that
  redshift. However, the  currently available data
  do not allow us to confirm if the galaxies lie at the same redshifts
as the absorption systems. A more extended structure is visible in the
SDSS~J14085+020522 field at an impact parameter of 0\farcs8 or
7~kpc. If these objects are at $z\approx2$ their luminosities are
0.03--0.04~$L^*$ in both fields. The star formation rates estimated from the
UV flux are 0.5--0.6~$M_{\odot}$~yr$^{-1}$, while the SFRs are half
these values if the $U$ band flux is due to \lya\ emission alone.
}
  % conclusions (optional)
{The non-detection of galaxies near to the line of sight is most likely
  explained by low metallicities and luminosities of the \ion{Mg}{ii}
  galaxies. Alternatively, the \ion{Mg}{ii} clouds are part of
  extended halos or in outflows from low-metallicity galaxies. }


   \maketitle
%
%________________________________________________________________

\section{Introduction}
During the past decade much information has been gathered about the
properties of high redshift galaxies from the surveys of Lyman break
galaxies \citep[LBGs;][]{steidel95,steidel03}. Since their detection
requires the galaxies to have relatively bright continuum emission,
the Lyman break technique preferentially selects massive galaxies
\citep{erb06a}, and follow up spectroscopy have revealed relatively
metal rich galaxies \citep[0.4--0.8 solar in][]{erb06b}.
Alternatively, the spectra of high redshift quasars can reveal the
existence of much fainter and more metal poor galaxies through
intervening absorption lines. Strong \ion{H}{i} absorption lines with
column densities in the damped Lyman-$\alpha$ (DLA) regime (log
$N$(\ion{H}{i}) (cm$^{-2}$)~$\geq$~20.3) are believed to arise when
the sight line toward a QSO intersects a gas rich galaxy
\citep{wolfe86,wolfe05}.

Metal absorption lines indicate typical DLA metallicities between one
hundredth and one tenth solar \citep{pettini99,prochaska02}. The
velocity profiles of the lines \citep{prochaska97} can be explained by
the complex dynamics of infalling clumps in a merging scenario
\citep{haehnelt98,ledoux98,nagamine07}. Numerical simulations
reproduce reasonably well the kinematics for $z=3$ DLAs in a
hierarchical model \citep{pontzen08}, where gas is later distributed
in discs at $z=0$.  Comparisons of local \ion{H}{i} discs with DLAs at
$z>2$ have indicated that local \ion{H}{i} discs have different
kinematics than high redshift DLAs \citep{zwaan08}, which may indicate
that some DLAs could arise in starburst winds or debris from tidal
interactions. Winds may not be solely responsible for the large
velocities indicated by the absorption lines. Observations have
indicated a relation between the velocity width of the metal
absorption lines and DLA metallicity \citep{ledoux06}, which has been
reproduced by numerical simulations \citep{pontzen08}. In one DLA
towards \object{Q0458-020}, the velocity difference between the
absorption lines and the \lya\ emission line seen in the DLA trough is
consistent with a rotating disc \citep{heinmuller06}. In most cases,
no \lya\ emission lines are found in the DLA trough, so whether the
star formation activity and supernovae that produce the metals occur
in situ \citep{wolfe04} at the time of the spectroscopic observation
of the DLA, still has to be verified observationally. High redshift
DLA systems might also be affected by galactic winds that remove the
neutral gas from the galaxies in competition with accretion from the
intergalactic medium \citep{bond01,prochaska09,tescari09}. Combining
measurements of star formation rates (SFRs) and impact parameters for
the galaxies with metallicities and velocities determined from the
absorption lines, is needed in order to disentangle these effects.

In contrast to LBGs, the galaxy counterparts of DLAs are selected
independently of their luminosities.  To date $\sim$1000 DLA systems
are known from the SDSS spectra \citep{prochaska05,noterdaeme09}. Yet,
despite intense observational efforts in the past couple of decades,
only four $z>2$ DLA galaxies have been found in emission
\citep{moller02,moller04}. The main difficulty is detecting the
galaxies against the glare of the bright background QSO, where the
continuum emission from the galaxies may be 10 magnitudes fainter than
the background source. High resolution images from the HST have
revealed faint ($H_{\mathrm{AB}}\sim25$) objects typically
1--2\arcsec\ in projection from the QSOs studied \citep{warren01}, but
without redshift information, it is not known whether these are the
galaxy counterparts to the DLAs.  Other techniques exploit the total
absorption of the QSO emission at the wavelength of \lya\ in the
DLA. If the absorbing galaxies are forming stars, they should be
detectable as \lya\ emitters and through the UV continuum emission
from their young stellar populations.  Narrow band images have
provided a few detections of \lya\ emission
\citep{smith89,wolfe92,moller93} as well as a few upper limits
\citep{grove09} close to the sight lines of QSOs.  Some candidate
\lya\ emitters are found at the redshift of the DLAs with integral
field spectroscopy \citep{christensen07}, while strong limits on the
\lya\ flux from DLA galaxies have been obtained with Fabry-Perot
images \citep{kulkarni06}.  DLA galaxies are fainter and have smaller
SFRs than typical LBGs \citep{fynbo99,colbert02,wolfe06,fynbo08}.
Very deep spectroscopic observations of \lya\ emission from candidate
DLA galaxies gives typical SFRs of a few tenths of solar masses per
year where the \lya\ emission can be extended over a few tens of kpc
\citep{rauch08}. The DLA galaxies are therefore challenging to detect
even with the largest telescopes.

Another method to detect the galaxies in emission takes advantage of
sight lines with multiple intervening strong absorbers. If the highest
redshift absorber has a hydrogen column density much larger than
10$^{17}$~cm$^{-2}$, defining a Lyman limit system (LLS), all the flux
from the QSO bluewards of the Lyman break ($\lambda=912(1+z)~{\AA}$) is
absorbed. This idea was originally exploited in the search for high
redshift galaxies towards luminous QSOs \citep{steidel92}.  If the
sight line has a lower redshift absorber, the emission from its
associated galaxy will be visible even directly in front of the QSO,
or at a very small impact parameter. \citet{omeara06} observed strong
\ion{Mg}{ii} absorbers at $z\sim2$ seen in QSO spectra which had
higher redshift LLS, and found two bright $L^*$ galaxies at projected
distances of 12--16 kpc to the two QSOs.  Several studies have aimed
to identify the host galaxies of strong \ion{Mg}{ii} systems, mostly
at lower redshifts $z\lesssim1$. The presence of galaxies responsible
for absorption lines at impact parameters of 20--40~kpc is used to
argue for starburst driven winds \citep{nestor07,bouche07}, but such
winds are not necessary in a model where the extended gaseous halos
follow a Holmberg relation \citep{steidel95b,kacprzak08,chen08}.

Gamma-ray burst (GRB) afterglows have recently proven to be of similar
use as QSOs for the study of intervening absorption systems. GRBs have
the advantage that the afterglows fade away providing a clear line of
sight for the absorbing galaxies. In a several cases, galaxies which
are possibly responsible for intervening strong \ion{Mg}{ii}
absorption lines have been found within an impact parameter of
$\sim$10~kpc from the line of sight to the afterglow
\citep{jakobsson04,chen09,pollack09}. These intervening galaxies have
luminosities in the range 0.1--1~$L^*$. The discrepancy between the
impact parameters and luminosities found for galaxy counterparts of
QSO-- and GRB \ion{Mg}{ii} absorbers could suggest that bright
background QSOs prevent us from detecting potentially fainter galaxies
closer to their lines of sight.

In this paper we exploit the absorption of the UV light of two high
redshift QSOs that have multiple intervening absorption systems in
their spectra.  The observations in the two fields reach an
unprecedented depth with a detection limit of 0.03$L^*$ in the $U$
band. This allows us to put strong constraints on the SFRs directly in
the QSO lines of sight.  Throughout the paper, we assume a flat
cosmology with $\Omega_{M}=0.3$, $\Omega_{\Lambda}=0.7$ and
$H_0$~=~72~km~s$^{-1}$~Mpc$^{-1}$.

%__________________________________________________________________

\section{Observations}

\subsection{Sample selection}

While Lyman limit systems with \ion{H}{i} column densities of
$10^{17}$~cm$^{-2}$ are optically thick bluewards of the Lyman limit,
some of the background emission is transmitted, and only at higher
column densities where the optical depth is sufficient
($N$(\ion{H}{i})$>10^{19}$~cm$^{-2}$) will the blue wavelengths be
completely absorbed. A QSO with a DLA system at $z>3.39$ will have all
its emission absorbed in the $U$ band. An intervening galaxy with a
redshift in the range $1.7<z<2.3$ will have a \lya\ emission line in
the $U$ band provided it is forming stars. Meanwhile, the faint
emission will be observable from the ground and its emission
properties will be completely unaffected by the bright background QSO.
This effective approach to locate the galaxies responsible for
intervening strong absorption lines (including DLAs), near bright
QSOs is demonstrated in \citet{omeara06}.

\begin{figure*}[htbp!]
\centering
  \resizebox{\textwidth}{!}{\includegraphics[bb=90 370 1020 700, clip]{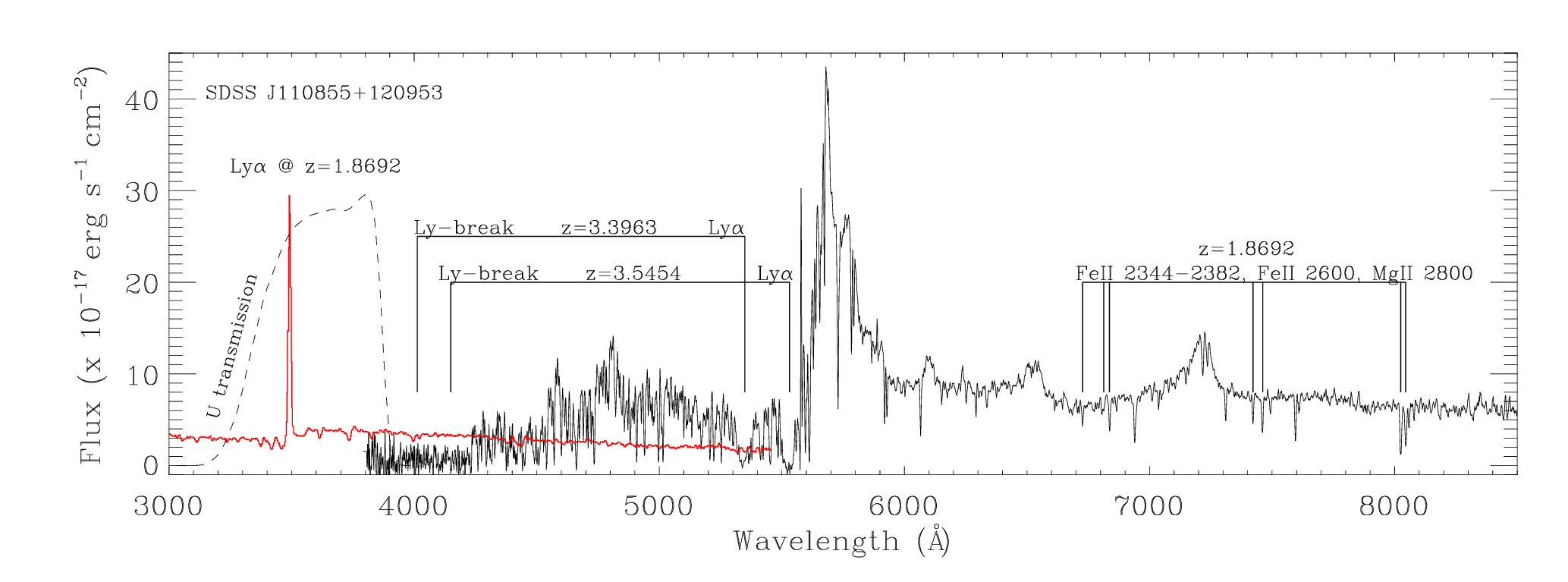}}
  \resizebox{\textwidth}{!}{\includegraphics[bb=90 370 1020 700, clip]{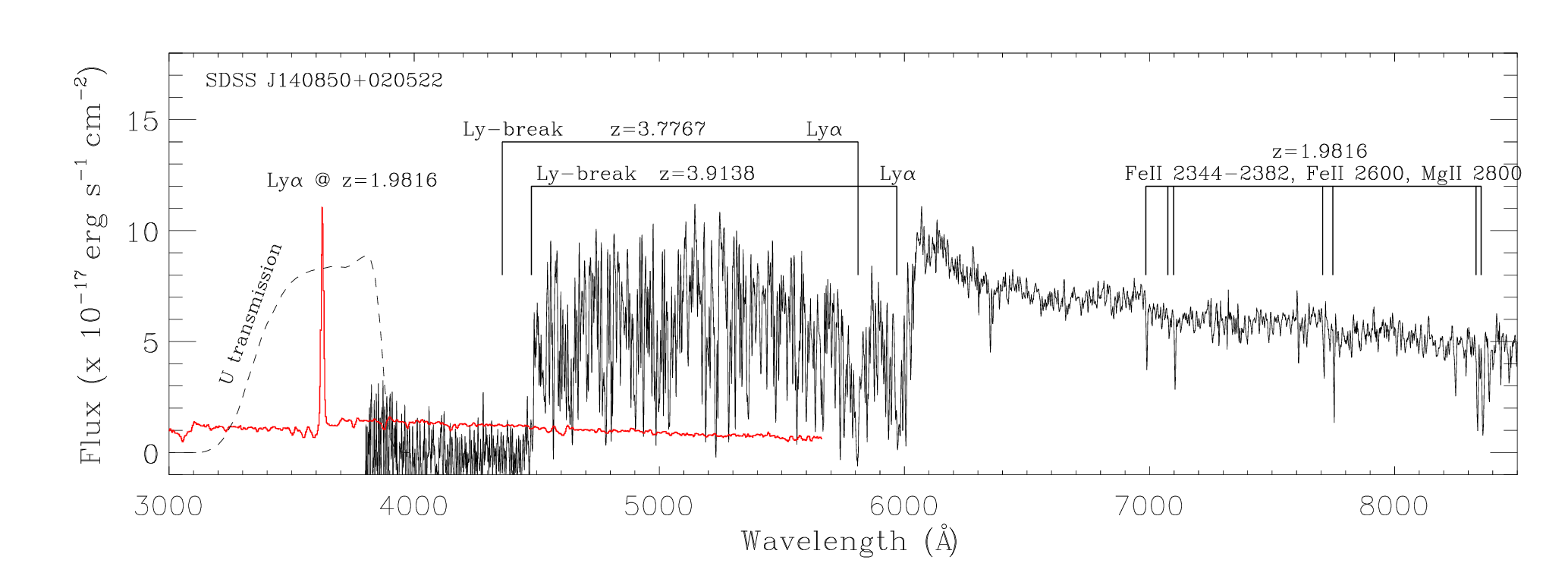}}
  \caption{Sloan spectra of the two quasars. In each quasar spectrum,
    two DLAs (at $z_{\mathrm{DLA}}$ = 3.5454 and 3.3963 for
    SDSS~J110855.46+120953.3, upper panel, and $z_{\mathrm{DLA}}$ =
    3.9138 and 3.7767 for SDSS~J140850.91+020522.7, lower panel)
    completely absorb the QSO emission bluewards of 4200~{\AA}. Two
    lower redshift systems are identified through their strong metal
    absorption lines. Both panels indicate the transmission curve of
    the $U$ band, which lies bluewards of the absorption cutoffs.
    Spectra of Lyman break galaxies with \lya\ in emission
    \citep{shapley03} at the redshifts of the \ion{Mg}{ii} absorption
    systems are indicated in the two plots. Their \lya\ emission lines
    fall within the $U$ band in both cases.}
   \label{fig:spectra}
\end{figure*}

To find such configurations, we systematically searched the Sloan
Digital Sky Survey spectra of about 2\,000 QSOs for $z>3.5$ DLAs in
the data release 5 \citep{schneider07}. To ensure that no flux was
present from the QSO and that the \lya\ emission line was well within
the transmission function of the $U$ band, the redshift criteria were
chosen to be conservative.  Because the QSO emission around the
\lya\ line of lower redshift galaxies is completely absorbed it is not
possible to determine the hydrogen column density. Instead we used the
strong \ion{Mg}{ii} $\lambda\lambda2796,2803$ doublet as a proxy for
selecting DLAs. These strong lines are easy to identify in QSO
spectra, and can be used to identify low redshift DLAs where only few
have measured hydrogen column densities \citep{rao00,rao06}. It is not
known if strong \ion{Mg}{ii} absorbers are identical to DLAs;
absorbers with \ion{Mg}{ii} equivalent widths larger than 0.6~{\AA}
and lines spread over more than 300~km~s$^{-1}$ only have a
probability of $\sim$50\% of being DLAs \citep{ellison09}. SDSS
Spectra with at least one high redshift DLA system were investigated
further to find \ion{Mg}{ii}~$\lambda\lambda$2796,2803 absorption
lines with a rest frame equivalent width ($W_{\mathrm{r}}$) larger
than 0.6~{\AA} at $1.75<z<2.2$, where the redshift limits ensure that
the associated \lya\ emission line fall within the FORS $U$ band.
These criteria were met by two equatorial QSOs
(\object{SDSS~J110855+120953} at $z=3.671$ and
\object{SDSS~J140850+020522} at $z=4.008$) which could be observed by
the VLT. Hereafter, we shall refer to these two QSOs as Q1108, and
Q1408. The SDSS spectra of the two QSOs are shown in
Fig.~\ref{fig:spectra}.  By coincidence both targets had two
intervening DLAs in addition to the strong \ion{Mg}{ii} system, which
effectively absorb all of the QSO emission immediately bluewards of
the redshifted Lyman limit break.

In each spectrum, the lower redshift absorption system is detected by
several metal absorption lines in addition to the
\ion{Mg}{ii}~$\lambda\lambda$2796,2803 doublet.
Table~\ref{tab:abssys} lists the properties of the absorption systems.
The high $W_{\mathrm{r}}$ measured for the metal lines suggest that
the strong metal absorption systems are possibly DLAs. At
$W_{\mathrm{r}}^{\ion{Mg}{i}\lambda2852}>$~0.6~{\AA}, combined with a
criterion for the fraction
$W_{\mathrm{r}}^{\ion{Mg}{ii}\lambda2796}/W_{\mathrm{r}}^{\ion{Fe}{ii}\lambda2600}<2$,
more than 50\% of the systems are DLAs \citep{rao06}. This is
consistent with the analysis of the \ion{Mg}{ii} absorption line
velocity spread discussed in Sect.~\ref{sect:velocity}.

\begin{table*}
\centering
\begin{tabular}{llll|lccc}
\hline\hline
  \noalign{\smallskip}
    & $z_{\mathrm{QSO}}$  & $z_{\mathrm{DLA,1}}$ &  $z_{\mathrm{DLA,2}}$ &  $z_{\mathrm{\ion{Mg}{ii}}}$ &
  $W_{\mathrm{r}}^{\ion{Fe}{ii}\lambda2600}$ &  $W_{\mathrm{r}}^{\ion{Mg}{ii}\lambda2796}$  &
  $W_{\mathrm{r}}^{\ion{Mg}{i}\lambda2852}$\\
      & & & & & ({\AA}) & ({\AA}) & ({\AA})\\
  \noalign{\smallskip}
\hline
   \noalign{\smallskip}
Q1108  & 3.671 & 3.5454 & 3.3963 & 1.8692 & 1.23 & 2.46 & 0.61\\  
Q1408  & 4.008 & 3.9138 & 3.7767 & 1.9816 & 0.97 & 1.89 & \\
  \noalign{\smallskip}
\hline
\end{tabular}
\caption{Absorption line systems in the two QSO spectra. The metal
  line rest frame $W_{\mathrm{r}}$ are measured
  for the lowest redshift systems in the SDSS spectrum for Q1408, and
  the UVES data for Q1108. The SDSS spectra give values consistent
  within the uncertainties with the UVES data. The SDSS spectrum of
  Q1408 is too noisy to identify the \ion{Mg}{i}
  line.  }
\label{tab:abssys}
\end{table*}

\subsection{Photometry}
The two fields were observed with the blue sensitive CCD on VLT/FORS1
taking advantage of the high transmission of a new $U$ filter
($U$-high), which transmits almost 90\% at its peak sensitivity around
3800~{\AA}.  We obtained very deep observations in the $U$ band (4.6
and 5.4 hours on target, respectively, divided into jittered exposures
of 980~s), with shorter (450~s) $R$ band observations for both
fields. The $R$ band images helped to pinpoint the exact location of
the QSO with respect to the nearby galaxies. The observations were
done in service mode between March 6 and 9, 2008 under photometric
conditions and with seeing conditions between 0\farcs6 and 1\arcsec.

\begin{figure}
   \centering
   \resizebox{0.24\textwidth}{!}{\includegraphics[bb=65 398 326 1173, clip]{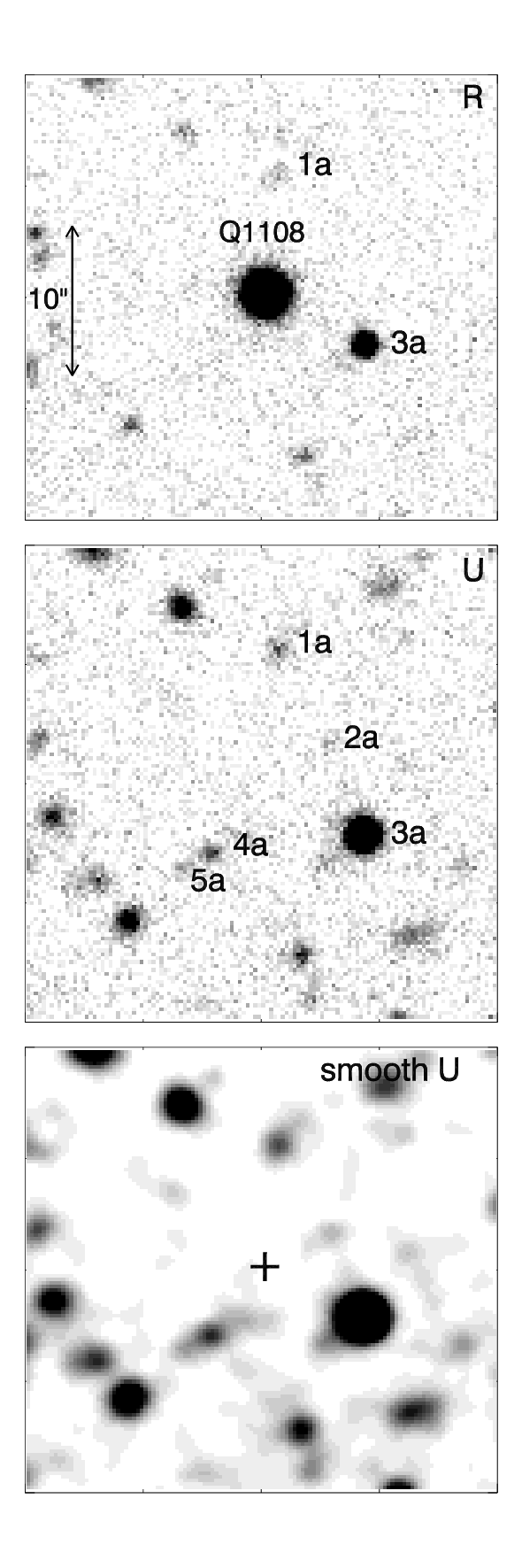}}
   \resizebox{0.24\textwidth}{!}{\includegraphics[bb=65 398 326 1173, clip]{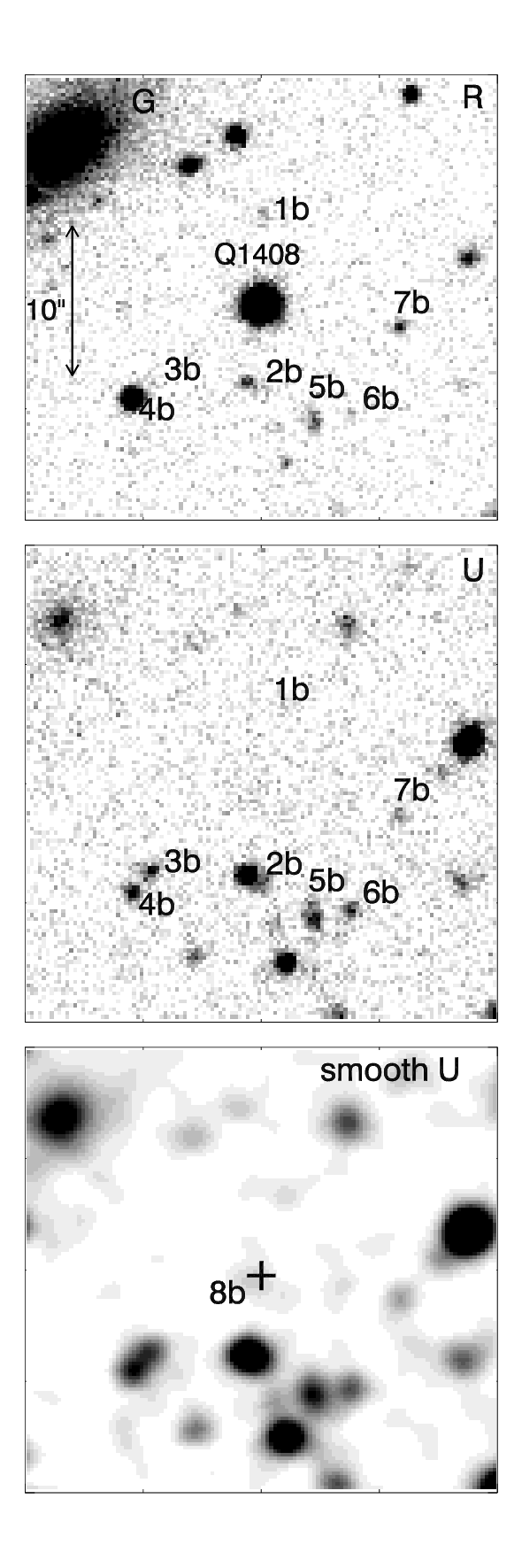}}
   \caption{$R$ and $U$ band images of Q1108 shown the left column,
     and for Q1408 in the right. In the $U$ bands the emission from
     the QSOs have been completely absorbed. The images are
     30\arcsec\ on a side with the orientation north up and east
     left. Objects within 10\arcsec\ from the QSOs lines of sight are
     labeled with numbers and the photometry of these is listed in
     Table~\ref{tab:maggies}. The bottom row shows smoothed version of
     the $U$ band images, with a `+' sign indicating the location of
     the QSO. Only in the case of Q1408 do we detect an extended
     object close to the line of sight of the QSO, which is otherwise
     not detectable in the unsmoothed $U$ band.}
   \label{fig:img1}
   \end{figure}

The data were reduced with standard procedures, subtracting an average
bias frame, and flat fielding using twilight sky frames. The images
were registered and combined to reject bad pixels and cosmic ray hits.
To optimise the depth of the images in the $U$ band, two of the 16
frames of Q1408 were rejected due to poor seeing.  The \textit{FWHM}
measured from stars in the fields is 0\farcs8 and 0\farcs9 in the $R$
and $U$ band, respectively.  Both $U$ band images do not show any
residual emission from the QSOs. The full field of view of FORS1 is
6\farcm8$\times$6\farcm8, but here we focus on the immediate regions
around the QSOs, as shown in Fig.~\ref{fig:img1}.  The images are
30\arcsec\ on a side with orientation north up and east left.

Instrumental zero points were estimated from observations of standard
stars in the field Rubin~149 obtained on the same nights as the
science observations. Transformations from instrumental magnitudes to
Vega magnitudes were calculated using {\tt IRAF-PHOTCAL}. The
transformation equations from the instrumental to Vega magnitudes
included an extinction coefficient appropriate for Paranal
\citep{patat03}. A potential inclusion of a colour term was consistent
with zero to within the uncertainties, so this term was neglected.
Magnitude uncertainties were propagated through the equations.
Aperture photometry showed that within a 1\arcsec\ radial aperture,
the 3$\sigma$ significance detection limit was
$U_{\mathrm{Vega}}$=27.3 mag for a point source in both fields, while
the 3$\sigma$ significance limit in the $R$ band was shallower: 26.4
mag and 26.5 mag (Vega) for Q1108, and Q1408, respectively. The
magnitudes were corrected for Galactic extinction \citep{schlegel98}.
Transformations from the Vega to AB magnitudes were calculated from
the filter transmission curve and the spectrum of Vega
\citep{fukugita95}. For the FORS1 filters we calculated
\(U_{\mathrm{AB}}=U_{\mathrm{Vega}}+0.66\) and
  \(R_{\mathrm{AB}}=R_{\mathrm{Vega}}+0.18\) mag. We use the AB
    magnitudes unless otherwise stated.

Since the $U$ band images were the deepest they were used primarily to
detect objects. The locations were cross checked with object detection
in the $R$ band to avoid missing any $U$ band dropout. 

\subsection{High resolution spectra}

In addition to the imaging data we took advantage of one 4200~s
integration of Q1108 with VLT+UVES obtained as a part of another
observing programme (ID 080.A-0482, PI: Sebastian Lopez). The data
were used to study the velocity widths of the strong \ion{Mg}{ii}
absorption lines as described in Sect.~\ref{sect:velocity}.

Both the blue and red UVES spectrograph arms were used simultaneously
with standard dichroic settings, with central wavelength of 437 and
860 nm. The resulting wavelength coverage is 305 to 1042~nm with gaps
at 575--583~nm and 852--866~nm. The CCD pixels were binned by a factor
of $2\times2$ and the slit width adjusted to 1\arcsec. This yielded a
resolving power of 48\,000 under the seeing conditions of 1\arcsec.

The data were reduced using the ESO pipeline system, which allowed for
an accurate extraction of the object spectrum, while subtracting the
sky spectrum and removing cosmic ray hits and CCD defects at the same
time.  Wavelengths were finally shifted to the vacuum-heliocentric
rest frame. The signal-to-noise ratio in the reduced spectrum was
measured to be 15--20 per pixel.

%________________________________________________________________

\section{Observational results}

\subsection{Nearby galaxies and colours}
\label{sect:colours}
The goal of this study is to detect objects located directly along the
QSO line of sight.  In both fields the closest objects lie at a
separation of $\sim$5\arcsec\ from the QSO. If either of these galaxies
are responsible for the strong absorption line system seen in the QSO
spectra, the projected separation is $\sim$40~kpc at the redshift of
the absorbers. No other nearby objects are found at the 3$\sigma$
significance level (corresponding to 0.03~$L^*$).  At $z=2$ a luminous
$L^*$ galaxy has an absolute magnitude of
$U_{\mathrm{AB}}^*$=--21.9~mag \citep{gabasch04}, and an observed
magnitude of $U_{\mathrm{AB}}=24.0$~mag in the adopted cosmology. The
numbered objects in Fig.~\ref{fig:img1} mark all galaxies that are
located within 10\arcsec\ of the QSO; most of these objects are
fainter than $U$~=~24~mag. The photometry of these galaxies is listed
in Table~\ref{tab:maggies}.

We investigate whether the colours are similar to other galaxies
detected in galaxy surveys. For comparison we use photometry of
galaxies from the GOODS-MUSIC catalog \citep{grazian06} where
multiband photometry allows an accurate determination of the
photometric redshift. To estimate the $R_{\mathrm{AB}}$ magnitude of
the GOODS galaxies, we linearly interpolate the flux between the
measured F606W and F775W bands to calculate the flux at 6440~{\AA}.
We are mainly interested in the observed colours of galaxies around
$z=2$, where the HST bands correspond to restframes 2000 and
2580~{\AA}, respectively. Between wavelengths 1500--2800~{\AA} galaxy
spectra are mostly flat when measured in frequency units, $f_{\nu}$
\citep{kennicutt98}, so a simple interpolation to calculate the $R$
band magnitude is well justified. The transmission curves for the
VIMOS $U$ band used in the GOODS-MUSIC catalog is different from the
FORS1 high transmission $U$ filter. We calculate a transformation
between the two systems: \( U_{\mathrm{FORS1}} =
  U_{\mathrm{VIMOS}}+0.04~\mathrm{mag}\).

\begin{figure}
 \centering
 \resizebox{0.5\textwidth}{!}{\includegraphics[bb=95 370 680 850, clip]{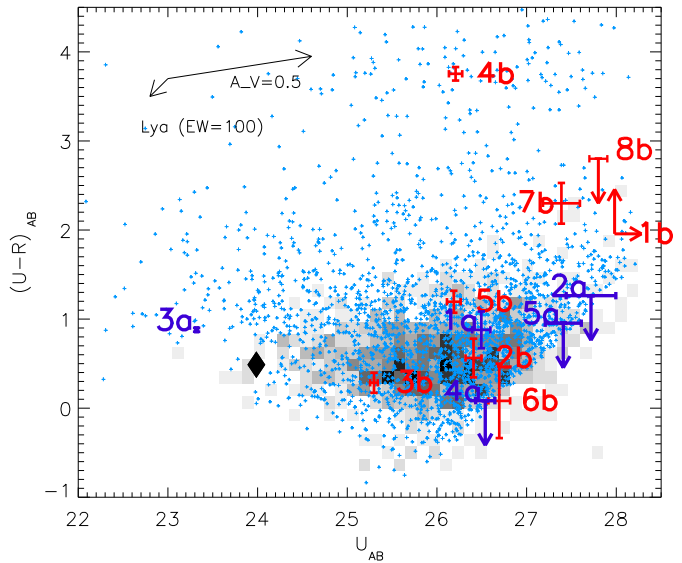}}
 \caption{Colour--magnitude diagram of the objects in the two fields
   (small plus signs). The grey scale image shows the distribution of
   $\sim1000$ galaxies with photometric redshifts
   $1.8<z_{\mathrm{phot}}<2.3$ from the GOODS-MUSIC survey. The large
   crosses with error bars mark the galaxies near to the QSO lines of
   sight and the numbers refer to those in Table~\ref{tab:maggies}
   (red symbols for Q1408, and purple for Q1108). The large black
   diamond symbol shows the colour of an $L^*$ starburst galaxy at
   $z=2$.  \textit{See the online edition of the journal for a colour
     version of this figure.}}
 \label{fig:cm}
 \end{figure}

A colour-magnitude diagram of $\sim$1000 galaxies with photometric
redshifts between 1.8 and 2.3 is shown as a grey scale image in
Fig.~\ref{fig:cm}. Galaxies detected in the two QSO fields with a
signal-to-noise ratio $>3$ are shown as small symbols, and the
objects detected within 10\arcsec\ are shown as larger symbols with
error bars. Due to our shallower $R$ band observations relative to the
GOODS observations, objects in the lower right hand corner are
undetected in our observations. The colours of the GOODS galaxies span
a narrower range than for the galaxies in the two QSO fields, where
there is an excess of galaxies with colours around $U-R>1.5$. If no
photometric redshift selection is made for the GOODS galaxies, a wider
range of colours are found, and the two colour distributions are
consistent.

The typical colour of an irregular galaxy, calculated from template
spectra \citep{kinney96}, and redshifted to $z=2$ is shown as a
diamond, where its $U$ band magnitude correspond to an $L^*$
galaxy. By creating artificial spectra of metal poor, young galaxies
\citep{bruzual03} with an additional intrinsic extinction of
$A_V$=0.5~mag, we estimate the colour and magnitude change. We also
calculate the colour and magnitude change if the galaxies have strong
\lya\ emission with a rest frame equivalent width of 100~{\AA}. The
arrows in Fig.~\ref{fig:cm} show the respective changes which are
minor.  More evolved galaxies have redder colours ($U-R>3$)
independently of the metallicity and the initial mass function of the
galaxy templates. This implies that the entire range of colours in
Fig.~\ref{fig:cm} can be obtained for high redshift galaxies.
Specifically, the red colour of object no. 4b in the Q1408 field is
consistent with an evolved galaxy at $z\approx2$ with an age greater
than 700~Myr for the dominant population of stars. In the Q1408 field
there are several red galaxies present at impact parameters larger
than 10\arcsec\ from the QSO line of sight. In the upper right hand
panel of Fig.~\ref{fig:img1}, the $R$ band image of Q1408, a large and
bright galaxy is partly visible (marked `G') in the image. This galaxy
has the morphology and colours of a low redshift elliptical, and there
are several fainter galaxies with the same colours around it. It is
likely that object no. 4b is part of a lower redshift group
environment.

Although there are no clear correlations between the $U-R$ colour and
redshift, objects with $U-R<0.5$ are more likely to lie at
$z>0.9$. 52\% of the GOODS galaxies fulfill these criteria, while only
20\% of galaxies at $z<0.9$ have such blue colours.  However, the
single $U-R$ colour is insufficient to make an exact redshift estimate, and
not good at all to select objects at $z\sim2$. Additional observations
in the $V$ band would be useful to apply the BM/BX colour criterion to
select galaxies with $1.4<z<2$ \citep{adelberger04}.

Objects no. 2b, 3b and 6b in the Q1408 field and no. 4a in the Q1108 field
have the bluest colours. Given their non-detection in the $R$ band,
several of the objects could potentially be located at higher
redshifts with very blue colours.

\begin{table*}
\centering      
\begin{tabular}{lllrlll}
  \noalign{\smallskip}
\hline\hline
  \noalign{\smallskip}
   Object No. & \multicolumn{2}{c}{offset (\arcsec)} & $b$ (\arcsec) & 
$U$ (mag) & $R$ (mag) & $L_U/L_U^*$ at $z=2$ \\
  \noalign{\smallskip}
\hline
  \noalign{\smallskip}
Q1108 \\               
1a&  0.7 E &8.4 N & 8.4 & 25.83$\pm$0.10 &   25.41$\pm$0.18 & 0.10\\%no823
2a&  4.1 W &2.4 N & 4.7 & 27.06$\pm$0.28 &$>$26.4           & 0.03\\%no1
3a&  6.4 W &3.4 S & 7.2 & 22.66$\pm$0.01 &   22.23$\pm$0.02 & 1.87\\%no722
4a&  3.4 E &4.8 S & 5.8 & 25.88$\pm$0.11 &$>$26.4           & 0.10\\%no713 
5a&  5.3 E &5.2 S & 7.4 & 26.75$\pm$0.28 &$>$26.4           & 0.03\\%no701
\hline
Q1408\\
1b&  0.0 E &6.4 N & 6.4 & $>$27.3        &   25.82$\pm$0.25 & $<$0.03\\ %no1   
2b&  1.0 W &5.4 S & 5.5 & 24.64$\pm$0.04 &   24.81$\pm$0.10 & 0.31\\ %no1027
3b&  6.9 E &5.0 S & 8.5 & 25.75$\pm$0.09 &   25.64$\pm$0.20 & 0.11\\ %no1029
4b&  8.2 E &6.4 S & 10.4& 25.55$\pm$0.07 &   22.25$\pm$0.02 & 0.13\\ %no1014

5b&  3.3 W &8.2 S & 8.9 & 25.53$\pm$0.08 &   24.79$\pm$0.10 & 0.13\\ %no995 
6b&  5.6 W &7.7 S & 9.5 & 26.03$\pm$0.12 &   26.41$\pm$0.40 & 0.08\\ %no1001
7b&  8.8 W &1.7 S & 8.9 & 26.73$\pm$0.21 &   24.88$\pm$0.10 & 0.04\\ %no1071
8b&  0.7 E &0.3 S & 0.8 & 26.92$\pm$0.10 &      --          & 0.04\\ % extended obj
  \noalign{\smallskip}       
\hline
\end{tabular}
\caption{Vega magnitudes of objects identified within
  $\sim$10\arcsec\ from the QSO lines of sight, with the observed
  offsets in RA and DEC (in arcsec), and projected impact parameter
  (in arcsec). Objects near Q1108 are in the upper 5 rows, and near
  Q1408 in the bottom 8 rows. The magnitudes are corrected for Galactic
  extinction. The last column gives the luminosity fraction relative
  to a $U^*$ galaxy at $z=2$, which has an apparent magnitude
  $U_{\mathrm{A}}=24$. The uncertainties are 0.01 for this fraction. }
\label{tab:maggies}
\end{table*}

\subsection{Extended objects}

To check for the presence of extended low surface brightness objects
closer to the lines of sight of the QSOs, the $U$ band images are
convolved with Gaussian point spread functions with a FWHM of 1\farcs5
and 2\arcsec, respectively. While the 3$\sigma$ detection limit is
$U_{\mathrm{AB}}=28.0$ for a point source, the smoothed images have a
deeper 3$\sigma$ detection limit of $U_{\mathrm{AB}}=28.5$ within a
1\farcs5 radial aperture for the two fields.  No objects are detected
close to the line of sight of the QSO in the field of Q1108. Objects
4a and 5a in the Q1108 field have a separation of 2\arcsec, and appear
merged into one elongated object in the smoothed image. Furthermore
there is an extension of emission to the west of object no.~4a which
is not directly visible in the original image. This extension reaches
a minimum impact parameter of 3\farcs8 directly south of the QSO with
a size of 1\farcs5$\times$3\arcsec\ and has a magnitude of
$U_{\mathrm{AB}}=27.4\pm0.1$. If the object is at $z\approx2$, the
projected distance is 31 kpc.

In the Q1408 field, a faint object ($U_{\mathrm{AB}}\sim27.6\pm0.1$ mag
within a radial aperture of 1\farcs5)
 is detected at an offset of 0\farcs8 to the south-east of
the QSO as displayed in Fig.~\ref{fig:img1}, and labeled as object
no. 8b. At $z=2$ the projected distance is 7 kpc. The object appears
elongated with a size of about 2\farcs5$\times$2\arcsec\ in the frame
with a Gaussian convolution width of 1\farcs5. To estimate whether a
faint object is present also in the $R$ band, the QSO is subtracted
using the brighter field stars as PSF reference. No additional point
sources were visible after the subtraction.  To estimate the detection
limit of a resolved object against the glare of the brighter QSO, we
add artificial objects nearby the bright QSO, and then subtract the
PSF to recover the artificial object. These experiments show that we
can recover an object with a \textit{FWHM} of 1\farcs5 and $R=25$~mag
at a distance of 0\farcs8 by PSF subtraction.  This places a rough
upper limit on the colour: $U-R<2.8$  for the object no. 8b.

\subsection{High redshift DLA host galaxies?}

 In the previous sections we investigated whether some of the
 spatially nearby galaxies are the hosts of the $z=2$ \ion{Mg}{ii}
 absorption systems, but a related question is whether some of the
 galaxies could be the hosts of the $z>3$ DLA systems.

For both fields, the two high redshift DLAs absorb all emission
immediately bluewards of the Lyman break. This suggests that the
objects detected in the $U$ band images are not likely to be the host
galaxies of the $z>3$ DLA systems because no UV emission bluewards of
the Lyman limit break should be able to escape the cloud, at least
along the QSOs line of sight.  Only object no. 1b in Q1408 is detected
in the $R$ band but not in the $U$ band, and this could potentially be
the host of one of the DLAs. However, the impact parameter is quite
large (44~kpc at $z\approx3.8$) compared to a typical value of
$\sim$10~kpc found for the few confirmed DLA galaxies at $z>2$
\citep{moller02,weatherley05}. Numerical models also predict small
impact parameters of the order or less than 10~kpc for DLA systems
\citep{nagamine07,pontzen08}. However, little is currently known about
the extension and morphology of gaseous disks around high redshift
galaxies from observations. 

Even though no UV photons bluewards of the Lyman limit can escape the
DLA clouds along the line-of-sight, they can escape along other
directions with a smaller optical depth. In such a special gas
geometry, the emission from the DLA host galaxies can be detected in
both the $U$ and the $R$ bands. If the escape fraction of the galaxies
is substantial, the Lyman continuum emission could be still observed,
and in the images the galaxy would be visible at some distance from
the QSO lines of sight.  Assuming an escape fraction
$f_{1500}/f_{900}$~=~2.9--4.5 as observed for two Lyman break galaxies
\citep{shapley06}, we can calculate the rest frame flux ratio which
corresponds to a colour $(U-R)_{\mathrm{AB}}\sim2.2$. Objects no. 4b
and 7b in the Q1408 field have colours consistent with being at a
redshift of $z>3$, but the colours are also consistent with lower
redshift evolved galaxies. With $R=22.25$ mag for object no. 4b, it
would have a luminosity of 7$L^*$ if it were at $z=3.8$ relative to an
$R^*=-23.0$ galaxy \citep{gabasch06} while 7b would have a luminosity
of 0.7$L^*$. Considering that such bright objects are rare, it is more
likely that object no. 4b has a lower redshift, and possibly is a
member of a group as discussed above.  In addition, object no. 7b
would have an impact parameter of 65~kpc at $z=3.8$. Since the objects
are faint for follow up spectroscopy, near-IR images of
the field can be used instead to investigate the spectral energy
distribution and to determine the photometric redshift. Before such
observations are made we cannot claim the detection of the hosts of
the DLAs.

%________________________________________________________________

\section{SFR limits}
\subsection{SFRs from the UV continuum}
\label{sect:sfrlim}
The limiting magnitude of $U_{\mathrm{AB}}=28.0$ corresponds to a SFR
of 0.6~M$_{\odot}$~yr$^{-1}$ for a galaxy at $z=2$. At the redshift of
the \ion{Mg}{ii} absorbers, the $U$ band measures the emission at
1200--1400~{\AA} in the rest frame of the absorbers. To estimate the
SFR in the UV region, we extrapolate calibrations from other studies.

The conventional conversion from a UV flux to the SFR
\citep{kennicutt98} is valid from 1500--2800 {\AA}, where spectra of
star forming galaxies are flat. Bluewards of 1500~{\AA}
the flux decreases, so we analyse template spectra to estimate the SFR
given the $U$ band flux. We use the template spectra from
\citet{bruzual03} of instantaneous star burst populations with ages
$<$100~Myr and a Salpeter initial mass function. Galaxy template
spectra are created with the observed $U$ band magnitude at the
redshift of the absorber. The template flux bluewards of 1215{\AA} is
reduced to reflect that 15\% of the flux is absorbed in the
\lya\ forest at $z=2$ \citep{daglio09}. No \lya\ emission is present
in the template spectrum. From this template the flux at the rest
frame 1500--2800~{\AA} is used to calculate the SFR using the standard
conversion \citep{kennicutt98}, which has an intrinsic scatter of
$\sim$30\%.

The SFR conversion depends also on the age and metallicity of the
template for which we have to select appropriate values.  Since strong
\ion{Mg}{ii} absorbers have about 50\% chance of being DLAs
\citep{rao06,ellison09}, we use typical low metallicities measured for
the DLA population \citep{wolfe05}.  Models of chemical evolutions of
the DLAs at high redshifts show that they are typically a few 100~Myr
old \citep{dessauges04}. A notable exception is the DLA towards
\object{Q~B2230+02}, which has a relatively high metallicity and an
age of 3~Gyr \citep{dessauges07}.  The conversion from luminosity to
SFR is given by

\begin{equation}
\mathrm{SFR}~(\mathrm{M}_{\odot}~\mathrm{yr}^{-1})  =  A \times10^{-28}
L_{\nu}~(\mathrm{erg~s}^{-1}~\mathrm{Hz}^{-1}),
\end{equation}
where $L_{\nu}$ is the luminosity measured around 1200~{\AA} in the
rest frame of the galaxy. For a 100~Myr, 0.2 solar metallicity
template, we calculate $A=2.9$ in Eq.~(1).  Since the factor $A$
depends on the metallicities and ages of the galaxies, we derive the
conversion factor for the SFR using a series of templates of varying
ages and metallicities (see Table~\ref{tab:conv}) corresponding to the
available template metallicities in \citet{bruzual03}.

If the \lya\ emission line is very strong we over estimate the SFR
with this procedure, since a pure \lya\ line indicates a smaller SFR
as described in Sect.~\ref{sect:lyaline}. Hence the SFRs that we give
are considered to be conservative.

\begin{table}
\centering
\begin{tabular}{lcccc}
  \noalign{\smallskip}
\hline\hline
  \noalign{\smallskip}
age &0.02$Z_{\odot}$ & 0.2$Z_{\odot}$ & 0.4$Z_{\odot}$ & 1$Z_{\odot}$ \\
  \noalign{\smallskip}
\hline
  \noalign{\smallskip}
10 Myr & 1.1 & 1.3 & 1.4 & 1.5\\
50 Myr & 1.7 & 1.9 & 2.3 & 2.9\\
100 Myr& 2.3 & 2.9 & 3.7 & 5.5\\
200 Myr& 3.5 & 5.1 & 7.3 & 14.9\\
  \noalign{\smallskip}
\hline
\end{tabular}
\caption{Conversion factor $A$ for Eq.~(1) calculated for different
  galaxy template ages and metallicities relative to the
  solar value, $Z_{\odot}=0.02$.}
\label{tab:conv}
\end{table}

It is possible that the galaxies responsible for the strong absorption
systems are extended low surface brightness objects.  We compute the
apparent magnitude and UV flux for an object detected at the 3$\sigma$
significance level as a function of radius.  
To calculate the limiting magnitude for a signal-to-noise ratio of
S/N=3 we use \(\mathrm{m_{lim}=zp} -2.5 \log (\mathrm{S/N}
  \sqrt{\mathrm{n_{pix}}} \sigma/t ) \), where zp is the FORS zero
 point measured in magnitudes, $\mathrm{n_{pix}}=\pi(r/0.25)^2$ is the
 number of pixels within an aperture radius $r$  (in arcsec), and
   0.25 is the plate scale of the CCD (\arcsec/pixel).  $\sigma$ is
 the standard deviation of the pixel values measured in the reduced
 CCD frame, and $t$ is the integration time  measured in
   seconds. Given these values, a point source has with a
   FWHM=1\arcsec\ has a radial $n_{\mathrm{pix}}=2$, which at $z=2$ 
corresponds to $\sim$4~kpc in the adopted cosmology.  The limiting 
flux increases with increasing aperture, so for more extended galaxies 
the SFR limit is less strong.

\begin{figure}
 \centering
 \resizebox{0.5\textwidth}{!}{\includegraphics[]{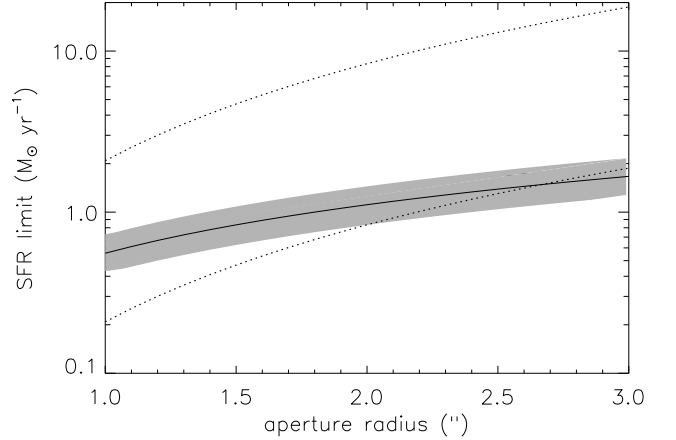}}
 \caption{The limiting SFRs in the two QSO lines of sight (solid line
   with the uncertainty represented as the shaded area) converted from
   the limiting magnitude calculated as a function of radial
   apertures.  The limits have been calculated from the original
   unsmoothed $U$ band images.  At $z=2$ one arcsec corresponds to
   8.1~kpc in projection. The upper dotted line shows the average SFR
   per unit area in DLA galaxies of
   $10^{-2}$~M$_{\odot}$~yr$^{-1}$~kpc$^{-2}$ while the lower dotted
   line represents a value of
   $10^{-3}$~M$_{\odot}$~yr$^{-1}$~kpc$^{-2}$ corresponding to the
   lower limit in \citet{wolfe04}.}
 \label{fig:sfrlim}
 \end{figure}

With the observations, we can place a strong constraint on the SFR for
an object located directly in the lines of sight of the QSOs.
Figure~\ref{fig:sfrlim} shows the limiting SFR for the lines of sight
to the two QSOs as a function of the radial aperture. At large
apertures of 3\arcsec, the SFR limit is within 2~M$_{\odot}$~yr$^{-1}$
including uncertainties.  Compared to typical LBGs which have an
average SFR measured from the unobscured UV emission of
8~M$_{\odot}$~yr$^{-1}$ \citep{erb06b}, our observations probe high
redshift galaxies which are significantly fainter.  Similarly,
extended continuum emission from DLA galaxies is not detected in the
Hubble ultra-deep field \citep{wolfe06}, possibly due to a smaller
star formation efficiency in DLAs relative to LBGs.

\citet{wolfe04,wolfe08} estimate the SFR surface density of DLA
galaxies to lie in the range $10^{-2}-10^{-3}$
M$_{\odot}$~yr$^{-1}$~kpc$^{-2}$ depending on the state of the neutral
gas. These values are represented by the dotted lines in
Fig.~\ref{fig:sfrlim}. If the absorbing galaxies we are looking for in
emission are similar to the DLA absorption systems investigated by
\citet{wolfe04} we should be able to detect them if the star formation
extends uniformly over the galaxy discs. Only the weakest
star-forming DLA galaxies would remain undetected. The non detections
indicate that galaxies with extended star formation, specifically at
the high rate of 10$^{-2}$~M$_{\odot}$yr$^{-1}$~kpc$^{-2}$, are
unlikely to be present in the two QSO lines of sight. Alternatively, a
high SFR density in a small dwarf galaxy system with a radius
$\lesssim$4~kpc has a SFR below the detection limit.

The most nearby point source objects (no. 2a,2b in the two fields)
have SFR=0.5~M$_{\odot}$~yr$^{-1}$ for Q1108 and
4.9~M$_{\odot}$~yr$^{-1}$ for Q1408, respectively, if they are the
absorbing galaxies. The extended object no. 8b near Q1408 has
SFR=0.6~M$_{\odot}$~yr$^{-1}$.

\subsection{SFRs from \lya\ emission}
\label{sect:lyaline}

In the hypothetical case that the $U$ band flux comes entirely from a
\lya\ emission line from the galaxies at $z=2$, the line flux for the
object no. 2a in Q1108 is
2.5$\times$10$^{-17}$~erg~cm$^{-2}$~s$^{-1}$. For Q1408 the line flux
is 10.0$\times$10$^{-17}$~erg~cm$^{-2}$~s$^{-1}$ and
1.3$\times$10$^{-17}$~erg~cm$^{-2}$~s$^{-1}$ for objects no. 2b and
8b, respectively. This in turns corresponds to a SFR of
0.5~M$_{\odot}$~yr$^{-1}$ for object no. 2a in the Q1108 field and
2.3~M$_{\odot}$~yr$^{-1}$ and 0.3~M$_{\odot}$~yr$^{-1}$ for objects
no. 2b and 8b near Q1408. To calculate these flux densities, we use
the case B recombination scenario \citep{osterbrock89}, a flux ratio
of \lya/\halpha~=~10, and the conversion between \halpha\ luminosity
and SFR from \citet{kennicutt98}.  This last assumption of a pure
\lya\ emission line object is not justified in the case of Q1408,
because the continuum emission is detected in the $R$ band, but it
serves the purpose of calculating the SFR for comparison to that
estimated from the UV continuum.

The \lya\ photons are effectively quenched in the presence of dust, so
the above SFR are strict lower limits. Since the SFR estimated from
the UV is higher than estimated from the \lya\ flux by a factor of 3,
the unobscured SFR may be somewhere between these values.

We note that the estimated \lya\ flux densities and luminosities are
in the same range as those measured for the few detections of
\lya\ emission from DLA galaxies \citep[][and references
  therein]{moller04,weatherley05}. This is probably not a coincidence,
but likely reflects that the searches for emission from DLA galaxies
are carried out to the detection limit of currently available
instruments.

%________________________________________________________________

\section{Strong \ion{Mg}{ii} systems}

\subsection{Velocity spreads}
\label{sect:velocity}
\begin{figure}
 \centering
 \resizebox{0.5\textwidth}{!}{\includegraphics[]{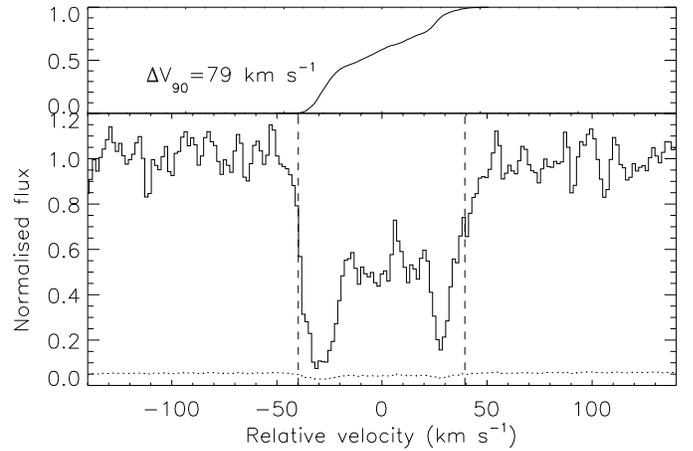}}
 \caption{The bottom panel shows the velocity profile of
   \ion{Fe}{ii} $\lambda2374$  from the UVES data, and the dotted
   line shows the error spectrum. The upper panel demonstrates the
   integrated, normalised, apparent optical depth, and the vertical dashed
   lines in the spectrum indicate the wavelengths within which 90\% of
   the absorption is seen. }
 \label{fig:vel}
\end{figure}

The UVES data of Q1108 shows a wealth of absorption lines and a
detailed analysis will be presented elsewhere. The systemic redshift
is found to be $z=1.8692$ from narrow metal absorption lines of
\ion{Mg}{i}~$\lambda$2852 and the \ion{Fe}{ii} lines.  The
\ion{Mg}{ii}~$\lambda\lambda$2796,2803 line profiles are complex with
several components spread over a velocity of $\Delta
v$~=~308~km~s$^{-1}$, as typically seen for strong \ion{Mg}{ii}
absorbers \citep{nestor07}. The same spread is seen in the \ion{C}{iv}
$\lambda$1549 doublet.  The bulk of the \ion{Mg}{ii} absorption is
spread over a smaller velocity range of $\approx$150~km~s$^{-1}$
surrounded by weaker satellites at larger velocities.  This places the
absorber among the strong, but not extremely strong \ion{Mg}{ii}
absorbers. A smaller velocity spread of the dominant absorption
components is seen from other non saturated lines such as the
\ion{Fe}{ii} $\lambda$2374 line, as demonstrated in
Fig.~\ref{fig:vel}. The velocity spread of the
\ion{Mg}{ii} line is twice that suggested by the correlation with the
equivalent width \( \Delta v~(\mathrm{km~s}^{-1}) \approx
  70~(\mathrm{km~s}^{-1}~\mathrm{{\AA}}^{-1}) \times
  W_{\mathrm{r}}^{\lambda2796}~(\mathrm{{\AA}})\)
 \citep{murphy07}. Using the \ion{Mg}{ii} absorption line velocity 
spread to calculate the $D$-index \citep{ellison06} gives 
$D=W_{\mathrm{r}}/\Delta v \times 1000=8.0$. For values of $D>7$ the 
probability that the absorber is a DLA is 50--55\% \citep{ellison09}.

To compare with previous studies \citep{ledoux06}, we measure the
velocity spread of the \ion{Fe}{ii} $\lambda$2374 line which satisfies
the criterion that the maximum residual emission lies between 0.1 and
0.6 times the continuum level. From the central wavelength defined by
the redshift of the system, we measure the velocity range over which
90\% of the apparent optical depth \citep{savage91} is seen.  We find
$\Delta V_{90}=79$~km~s$^{-1}$ as demonstrated in
Fig.~\ref{fig:vel}. Unsaturated lines typically have smaller velocity
spreads relative to the \ion{Mg}{ii} lines, which may be composed of
more components that contribute to the total line width due to their 
stronger transition.

\subsection{Metallicities}
\label{sect:metal}
Since the \lya\ absorption lines corresponding to the \ion{Mg}{ii}
systems lie in the absorbed parts of the QSO spectra, we cannot
determine the metallicities as usually done for strong absorption
systems. In the absence of an exact metallicity measurement for the
\ion{Mg}{ii} absorbers, we assume that the absorbing clouds are
actually DLAs. Consequently, some derivations in this section are only
valid if the objects are DLAs, for which the probability is 50\%.  In
case the \ion{Mg}{ii} absorbers are sub-DLAs we point out the
conclusions for this category.  We use calibrations and scaling
relations found for DLAs and sub-DLAs in the literature to derive the
metallicities, and use the conventional notation of the metallicity
[M/H]~$\equiv$~log[$N$(M)/$N$(H)]~--~log[$N$(M)/$N$(H)]$_{\odot}$.

A relation between the metallicity and velocity spread as measured by
the $\Delta V_{90}$ parameter has been found for DLAs around $z=2$
\citep{ledoux06}. Using this relation and the spread measured from the
UVES data of Q1108, the metallicity of the galaxy is expected to be
[M/H]~=~--1.4$\pm$0.4. A similar correlation exists for sub-DLAs, for
which \citet{dessauges09} suggest a [Fe/H] metallicity which is larger
by 0.4--0.6 dex for a given velocity spread. Using [Zn/H] as a tracer
of the sub-DLA metallicities, \citet{meiring09} find no correlation
with velocity, but find systematically larger values for sub-DLAs
relative to DLAs.

Another calibration uses the measured values of
$W_{\mathrm{r}}^{\ion{Si}{ii}\lambda 1526}$ which correlates with the
metallicity in high redshift DLAs \citep{prochaska08}.  To use this
relation for the \ion{Mg}{ii} absorber towards Q1108 we 
need  $W_{\mathrm{r}}^{\ion{Si}{ii}\lambda1526}$, but that
line is strongly contaminated by other absorption lines in the
\lya\ forest of the QSO. Instead we use the observed relation
which suggests that the widths of the lines roughly scale inversely
with their wavelengths:
$W_{\mathrm{r}}^{\ion{Mg}{ii}\lambda2796}$/$W_{\mathrm{r}}^{\ion{Si}{ii}\lambda1526}=3$
\citep{prochaska08}.  These relations suggest a metallicity
      [M/H]~=~--1.0$\pm$0.1 for the \ion{Mg}{ii} absorber in agreement
      with the previous metallicity estimate.

These low metallicities are typical for high redshift DLAs
\citep{pettini94,prochaska03b}. Observations have indicated that
sub-DLAs have higher metallicities on the average
\citep{peroux03,peroux08,meiring09}, while an investigation of the
redshift dependence demonstrated that this is only true at lower
redshifts ($z<1.7$) \citep{dessauges09}. 

\subsection{Expected galaxy magnitudes}
We can estimate the magnitude for low metallicity galaxies if a
luminosity metallicity relation exists for DLA galaxies
\citep{ledoux06}, again assuming that the clouds are DLAs. A DLA
galaxy with 10\% solar metallicity is expected to be faint:
$R\sim26.1$ mag or 0.06$L^*$ at $z=2$ according to the
metallicity--luminosity relation in \citet{ledoux06}. Such a relation
is valid provided the absorption lines velocity spread is used as a
proxy for the dark matter halo circular velocity, and there is 
a direct relation between the halo mass and a galaxy luminosity
\citep{haehnelt00}. In the case of a lower metallicity ([M/H]=--1.8)
the galaxy would be even fainter: $R\sim30$~mag.  Extrapolating this
relation to derive magnitudes for sub-DLAs with metallicities as large
as [M/H]=--0.4 gives $R=23.1$.  As described in
Sect.~\ref{sect:colours}, the $U-R$ colour of galaxies at $z=2$
depends on the age of the most recent dominant starburst, but
generally blue colours are expected for young galaxies. Thus, the
expected magnitude is $U\sim26.6$ in the 10\% solar metallicity
case. Our survey is deeper than this limit, so we would be able to
detect the continuum emission from the \ion{Mg}{ii} galaxies.  The non
detection of emission within 2\arcsec\ of line of sight is consistent
with the very low metallicity case. If the cloud is a sub-DLA instead,
a brighter host galaxy is possible, but since no bright galaxies are
found close to the QSO lines of sight, it would have to have a large
impact parameter.

Using instead the observed luminosity-metallicity relation either at
low \citep{tremonti04} or at higher redshift \citep{erb06b}, we rely
on extrapolation of the observed relations and metallicities derived
from emission line diagnostics.  We assume that metallicities
determined from emission and absorption lines are the same, and that
there are no metallicity gradients.  The luminosity metallicity
relation in \citet{tremonti04} shifts by 0.35~dex in metallicity
relative to the relation for galaxies at $z\sim2$ \citep{erb06b}.
This relation suggests that a 10\% solar metallicity DLA galaxy
has $U\sim31$~mag, i.e. well below the detection limit of our data,
while a 0.4 solar metallicity sub-DLA has $U\sim28$~mag.

For the strong \ion{Mg}{ii} absorber towards Q1408, we can only
estimate the metallicity from the
$W_{\mathrm{r}}^{\ion{Mg}{ii}\lambda2796}$ calibration in
\citet{turnshek05} and \citet{murphy07}.  This calibration gives
--0.9$<$[M/H]$<$--0.7, provided that the cloud is a DLA system.
While higher resolution spectra are needed to investigate the velocity
spread, the SDSS spectra indicate that the \ion{Mg}{ii} line of Q1408
has a width that is 20--30\% lower than in Q1108. Hence the
metallicity could be lower, and the absorbing galaxy
should be fainter than that towards Q1108.

%________________________________________________________________

\section{Discussion and summary}
Using the absorption of background QSO light by intervening DLAs at
$z>3$, we look for the galaxies responsible for intervening strong
\ion{Mg}{ii} systems at $z\approx2$ along two lines of sight. The
velocity spreads and equivalent widths of the absorption lines
indicate that the strong \ion{Mg}{ii} systems are possible DLAs. Very
deep images obtained for these two QSOs reveal no galaxies directly in
the line of sight to a limiting magnitude of $U_{\mathrm{AB}}=28.0$.
The most nearby objects are located at impact parameters of
$\sim$5\arcsec, corresponding to about 40~kpc at $z\approx2$. While no
point sources are found close to the lines of sight, we find evidence
for the presence of a more extended structure in a smoothed $U$ band
image of the field of Q1408. This structure has a much smaller impact
parameter of 0\farcs8 or 7~kpc at $z\approx2$.  Such extended low
surface brightness objects would be impossible to detect after PSF
subtraction of the bright background QSOs.

We consider two possible reasons for the non detections of nearby
point sources.  Either the galaxies are too faint, or the impact
parameters are large. The first possibility, that the galaxies are
very faint and below the detection limit, is in agreement with low
metallicity absorbers.  However, this hypothesis must be justified
based on several assumptions and extrapolations. 

Firstly, since the metallicities cannot be measured exactly because
the \lya\ absorption lines lie in the absorbed part of the QSO
spectra, we have to rely on the correlation between DLA/sub-DLA
metal-line velocity width and metallicity to estimate [M/H]. Secondly,
we must assume that the objects are DLAs, and not LLS which generally
have larger metallicities for a given absorption line width. Finally,
we must extrapolate the metallicity luminosity relation observed for
DLAs. These assumptions lead to an expected magnitude of the galaxy
below the detection limit.

The closest galaxies to the line of sight of the QSOs are located at
40~kpc.  They have luminosities of 0.03$L^*$ and 0.3$L^*$
corresponding to SFR of 0.5 and 4.9~M$_{\odot}$~yr$^{-1}$,
respectively for the Q1108 and Q1408 fields assuming that the
redshifts are indeed $z\approx2$. The luminosities are in agreement
with \citet{rao03}, who found that \ion{Mg}{ii} absorbers at $z<1$
arise in 0.1~$L^*$ galaxies, but in contrast to the results in
\citet{omeara06}, who found brighter galaxies (0.3--1.2$L^*$ within 25
kpc) for two different fields at $z\approx2$. In comparison, the
strong \ion{Mg}{ii} galaxies which intervene the sight lines to GRBs
are of similar luminosities than in the objects in the two FORS
fields, but are typically found at smaller impact parameters
\citep{pollack09}. The galaxies studied by \citet{omeara06} could,
however, be at lower redshift than the absorption
systems. Spectroscopic confirmation in these four fields including
ours would be of great interest.  The SFRs we derive for the two
fields are similar to those determined from spectroscopic observations
of the galaxies responsible for strong \ion{Mg}{ii} systems at lower
redshifts ($0.8<z<1$) \citep{bouche07}, but these galaxies are
generally found at smaller impact parameters (20$\pm$12~kpc).

While the impact parameters for the two closest objects are larger
than the size of neutral gas discs in high redshift gas rich
(proto)-galaxies as estimated in simulations \citep{nagamine07},
observational results in this area is still very limited. The
kinematics of high redshift DLAs are inconsistent with a large
rotating disc scenario \citep{zwaan08}. In order to explain the larger
velocities of high redshift DLAs relative to local \ion{H}{i} discs,
there may be a population of DLAs that arise in starburst winds or
from tidal interactions of galaxies, just as hypothesised for strong
\ion{Mg}{ii} systems \citep{bond01,bouche07}.  Clouds with DLA column
densities can be located several tens of kpc from the galaxy centre as
seen from observations \citep{ellison07} which is also supported by
simulations of the halos of massive galaxies at redshifts $z=3$
\citep{pontzen08}.

In order to investigate whether the \ion{Mg}{ii} absorption systems
are associated with the low luminosity galaxies at a considerable
impact parameter, spectroscopic data of the galaxies are necessary,
which is challenging due to their faintness. Nevertheless, the
\lya\ fluxes for pure emission line objects are within the reach of
current spectrographs. Specifically, IFU observations are useful in
the search for \lya\ emission lines very near to the QSO lines of
sight, and especially when the objects are extended. Such observations
can simultaneously be used to determine the redshifts of the other
galaxies within 10\arcsec\ of the line of sight to the QSOs.

\begin{acknowledgements}
We thank Sebastian Lopez for sharing the UVES data, and the referee
for a thoughtful and detailed report. PN acknowledges support from
the French Ministry of European and Foreign Affairs.

\end{acknowledgements}

\bibliography{ms12015}
\end{document}